\documentclass[12pt]{article}
\usepackage{amsmath}
\usepackage{amssymb} 
\usepackage[small]{caption2} 
\usepackage{fleqn} 
\usepackage{graphicx} 
\usepackage{mathrsfs}	
\usepackage[small,loose]{subfigure}  
\usepackage{cite} 
\advance \headheight by 3.0truept

\newcommand{\CenterObject}[1]{\ensuremath{\vcenter{\hbox{#1}}}}
\newcommand{\D}{\mathrm{d}}
\newcommand{\I}{\mathrm{i}}
\newcommand{\stau}{\widetilde{\tau}}
\renewcommand{\thesubsection}{\arabic{subsection}}

\begin{document}
\title{{\normalsize DESY 04-010\hfill\mbox{}\\
\normalsize UT 04-07\hfill\mbox{}\\
February 2004\hfill\mbox{}}\\
\vspace{1cm}
\textbf{Supergravity at Colliders}\\[8mm]}
\author{Wilfried Buchm\"uller$^{(a)}$, Koichi Hamaguchi$^{(a)}$, \\
Michael Ratz$^{(a)}$ and Tsutomu Yanagida$^{(b)(c)}$ \\[0.4cm]
{\normalsize\textit{$^{(a)}$Deutsches Elektronen-Synchrotron DESY, 
22603 Hamburg, Germany}}\\
{\normalsize\textit{$^{(b)}$Department of Physics, University of Tokyo, 
Tokyo 113-0033, Japan}}\\
{\normalsize\textit{$^{(c)}$Research Center for the Early Universe,  
University of Tokyo,
Japan}}
}
\date{}
\maketitle
\thispagestyle{empty}

\begin{abstract}
\noindent
We consider supersymmetric theories where the gravitino is the
lightest superparticle (LSP). Assuming that the long-lived
next-to-lightest superparticle (NSP) is a charged slepton, we 
investigate two complementary ways to prove the existence of
supergravity in nature. The first is based on the NSP lifetime 
which in supergravity depends only on the Planck scale and the NSP and
gravitino masses. With the gravitino mass inferred from
kinematics, the measurement of the NSP lifetime will test an unequivocal
prediction of supergravity. The second way makes use of the 3-body NSP decay.
The angular and energy distributions and the polarizations of the final state 
photon and lepton carry the information on the spin of the gravitino and
on its couplings to matter and radiation.
\end{abstract}

\newpage
\subsection{Introduction}\label{sec:Introduction}

Deciphering hidden symmetries in nature has been one of the most
exciting and challenging tasks in physics. Most recently, the 
discovery of the massive $W$ and $Z$ gauge bosons has established a spontaneously 
broken gauge symmetry as the basis of the electroweak theory.
Here, we discuss how one may discover the massive gravitino, which would
establish spontaneously broken local supersymmetry as a fundamental, hidden
symmetry of nature.

If the theory underlying the standard model is supersymmetric, one may find
superpartners of quarks, leptons and gauge bosons at the Tevatron,
the LHC or a future Linear Collider. Even though an exciting discovery, 
this would still not answer the
question how supersymmetry is realized in nature. To identify supersymmetry
as an exact, spontaneously broken symmetry requires evidence for the goldstino.
Only the discovery of the massive spin-3/2 gravitino, containing the spin-1/2 
goldstino, would establish supergravity \cite{Freedman:1976xh} with 
local supersymmetry as the fundamental structure.

In general it is difficult to detect gravitinos since their couplings are
Planck scale suppressed. However, evidence for the gravitino may be obtained in
collider experiments if it is the lightest superparticle (LSP). 
The gravitino mass may be of the same order as
other superparticle masses, like in gaugino mediation 
\cite{Kaplan:1999ac} or gravity mediation \cite{Nilles:1984ge}.
But it might also be much smaller as in gauge mediation scenarios 
\cite{Giudice:1998bp}. As we shall see, a discovery of the gravitino appears
feasible for gravitino masses in the range from about $1\,\mathrm{GeV}$ to
$100\,\mathrm{GeV}$. As LSP the gravitino is also a natural dark matter candidate.

We will assume that the next-to-lightest superparticle (NSP) is a charged 
slepton. This is a natural possibility with respect to the renormalization group
analysis of supersymmetry breaking parameters. Scalar leptons may be
produced at the Tevatron, the LHC or a Linear Collider. They can be directly
produced in pairs or in cascade decays of heavier superparticles. The NSP lifetime 
is generally large because of the small, Planck scale suppressed coupling to the 
gravitino LSP. 

The production of charged long-lived heavy particles at colliders is an exciting
possibility~\cite{Drees:1990yw}. Sufficiently slow, strongly ionizing sleptons 
will be stopped within 
the detector. One may also be able to collect faster sleptons in a storage ring.
In this way it may become possible to study NSP decays. The dominant
NSP decay channel is $\widetilde{\ell}\to\ell+\text{missing energy}$,
where $\widetilde{\ell}$ and $\ell$ denote slepton and lepton, respectively.

In the following we shall study how to identify the gravitino as cause of the
missing energy. First, one will measure the NSP lifetime. Since the gravitino
couplings are fixed by symmetry, the lifetime is predicted by supergravity
given the gravitino mass, which can be inferred from kinematics.
In a second step spin and couplings of the gravitino or the goldstino can be 
determined from an analysis of the 3-body decay 
$\widetilde{\ell}\to\ell+\psi_{3/2}+\gamma$.

\subsection{Gravitino mass}
\label{sec:GravitinoMass}

To be specific, we focus in the following on the case where the scalar  
lepton $\stau$, the superpartner of the $\tau$-lepton, is the NSP. 
It is straightforward to extend the discussion to the case where
another scalar lepton is the NSP. As we shall see, phenomenologically
particularly interesting is the case where the gravitino is not
ultra-light, which implies a long NSP lifetime.

At LHC one expects $\mathcal{O}(10^6)$ NSPs per year which are mainly
produced in cascade decays of squarks and gluinos \cite{Beenakker:1997ch}.  
The NSPs are mostly produced in the forward direction~\cite{Maki:1998ih}
which should make it easier to accumulate $\stau$s in a storage ring.  
In a Linear Collider an integrated luminosity of $500\,\mathrm{fb}^{-1}$
will yield $\mathcal{O}(10^5)$ $\stau$s \cite{Aguilar-Saavedra:2001rg}. Note that,
in a Linear Collider, one can tune the velocity of the produced $\stau$s by 
adjusting the $e^+e^-$ center-of-mass energy.

A detailed study of the possibilities to accumulate $\stau$ NSPs is beyond
the scope of this Letter. In the following we shall assume that a sufficiently
large number of $\stau$s can be produced and collected. Studying their decays 
will yield important information on the nature of the LSP. In the context of 
models with gauge mediated supersymmetry breaking the production of $\stau$ 
NSPs has previously studied for the Tevatron~\cite{Feng:1997zr},
for the LHC~\cite{Ambrosanio:2000ik} and for a Linear 
Collider~\cite{Ambrosanio:1999iu}.

The NSP $\stau$ is in general a linear combination of
$\stau_\mathrm{R}$ and $\stau_\mathrm{L}$, the superpartners
of the right-handed and left-handed $\tau$-leptons $\tau_\mathrm{R}$ and 
$\tau_\mathrm{L}$, respectively,
\begin{equation}\label{eq:StauLinearCombination}
 \stau\,=\,
 \cos(\varphi_\tau)\,\stau_\mathrm{R}+\sin(\varphi_\tau)\stau_\mathrm{L}\;.
\end{equation}
The interaction of the gravitino $\psi_{3/2}$ with scalar and fermionic
$\tau$-leptons is described by the lagrangian \cite{Wess:1992cp},
\begin{eqnarray}
 \mathscr{L}_{3/2}
 & = & 
 -\frac{1}{\sqrt{2}M_\mathrm{P}}
 \left[
   (D_\nu\,\stau_\mathrm{R})^*\overline{\psi^\mu}\,\gamma^\nu\,\gamma_\mu\,
   P_\mathrm{R} \tau
   +
   (D_\nu\,\stau_\mathrm{R})\,\overline{\tau} P_\mathrm{L}
   \gamma_\mu\,\gamma^\nu\,\psi^\mu\right]\;,
 \label{eq:FullGravitinoLagrangian}
\end{eqnarray}
where $D_\nu\,\stau_\mathrm{R} = (\partial_\nu + \I e\, A_\nu)
\stau_\mathrm{R}$. Here $A_\nu$ denotes the gauge boson, and
$M_\mathrm{P}=(8\pi\, G_\mathrm{N})^{-1/2}$ is the reduced Planck mass.  The
interaction lagrangian of $\stau_\mathrm{L}$ has an analogous form.

The $\stau$ decay rate is dominated by the two-body decay into $\tau$
and gravitino,
\begin{eqnarray}
 \Gamma_{\stau}^\mathrm{2-body}
 & = &
 \frac{\left( m_{\stau}^2 - m_{3/2}^2 - m_{\tau}^2 \right)^4 }{
	48\pi\,m_{3/2}^2\,M_\mathrm{P}^2\,m_{\stau}^3 }\,
 \left[1-\frac{4m_{3/2}^2\,m_{\tau}^2}{
 	\left( m_{\stau}^2 - m_{3/2}^2 - m_{\tau}^2 \right)^2} \right]^{3/2}\;,
 \label{eq:2bodyDecayRateWithTau}
\end{eqnarray}
where $m_\tau=1.78\,\mathrm{GeV}$ is the $\tau$ mass, $m_{\stau}$ is
the $\stau$ mass, and $m_{3/2}$ is the gravitino mass. Neglecting
$m_{\tau}$, we arrive at
\begin{equation}\label{eq:2bodydecay}
 \Gamma_{\stau}^\mathrm{2-body}
 \,=\,
 \frac{m_{\stau}^5}{48\pi\,m_{3/2}^2\,M_\mathrm{P}^2}
	\times\left(1-\frac{m_{3/2}^2}{m_{\stau}^2}\right)^4\;.
\end{equation}
For instance, $m_{\stau}=150\,\mathrm{GeV}$ would imply a lifetime of
$\Gamma_{\stau}^{-1}\simeq 78\,\mathrm{s}$ or
$\Gamma_{\stau}^{-1}\simeq 4.4\,\mathrm{y}$ for a gravitino mass of
$m_{3/2}=0.1\,\mathrm{GeV}$ or $m_{3/2}=75\,\mathrm{GeV}$,
respectively.
The crucial point is that the decay rate is completely determined by
the masses $m_{\stau}$ and $m_{3/2}$, independently of other SUSY
parameters, gauge and Yukawa couplings.  The mass $m_{\stau}$ of the
NSP will be measured in the process of accumulation.  Although the
outgoing gravitino is not directly measurable, its mass can also be
inferred kinematically unless it is too small,
\begin{eqnarray}
  m_{3/2}^2 &=& m_{\stau}^2 + m_\tau^2 -2m_{\stau} E_\tau\;.
\end{eqnarray}
Therefore, the gravitino mass can be determined with the same accuracy as
$E_\tau$ and $m_{\stau}$, i.e.\ with an uncertainty of a few GeV.

Comparing the decay rate \eqref{eq:2bodyDecayRateWithTau}, using the
kinematically determined $m_{3/2}$, with the observed decay rate, it
is possible to test an important supergravity prediction.  In other words,
one can determine the `supergravity Planck scale' from the NSP decay
rate which yields, up to $\mathcal{O}(\alpha)$ corrections,
\begin{eqnarray}
 M_\mathrm{P}^2(\text{supergravity}) & = &
 \frac{\left( m_{\stau}^2 - m_{3/2}^2 - m_{\tau}^2 \right)^4 }{
	48\pi\,m_{3/2}^2\,m_{\stau}^3\, \Gamma_{\stau}}\,
 \left[1-\frac{4m_{3/2}^2\,m_{\tau}^2}{
 	\left( m_{\stau}^2 - m_{3/2}^2 - m_{\tau}^2 \right)^2} \right]^{3/2}\;.
\end{eqnarray}
The result can be compared with the Planck scale of Einstein gravity,
i.e.\ Newton's constant determined by macroscopic measurements,
$G_\mathrm{N}=6.707(10)\cdot10^{-39}\,\mathrm{GeV}^{-2}$
\cite{Hagiwara:2002fs},
\begin{eqnarray}
  M_\mathrm{P}^2(\mathrm{gravity}) &=& (8\pi\, G_\mathrm{N})^{-1} 
  \,=\, (2.436(2)\cdot 10^{18}\,\mathrm{GeV})^2\;.
\end{eqnarray}
The consistency of the microscopic and macroscopic determinations of
the Planck scale is a crucial test of supergravity.

Furthermore, the measurement of the gravitino mass yields another
important quantity in supergravity, namely the mass scale of spontaneous
supersymmetry breaking,
\begin{equation}
 M_\mathrm{SUSY}\,=\,
 \sqrt{\sqrt{3}M_\mathrm{P}\,m_{3/2}}\;.
\end{equation} 
This is the analogue of the Higgs vacuum expectation value $v$ in the
electroweak theory, where $v = \sqrt{2}m_W/g = (2\sqrt{2}G_\mathrm{F})^{-1/2}$.

\subsection{Gravitino spin}\label{sec:GravitinoSpin}

If the measured decay rate and the kinematically determined mass of the 
invisible particle are consistent, we already have strong evidence for 
supergravity and the gravitino LSP. In this section we analyze how to
determine the second crucial observable, the spin of the invisible particle.

\begin{figure}[!h]
\begin{center}
 \subfigure[{}\label{fig:4particleInteractionM1R}]{
		\CenterObject{\includegraphics{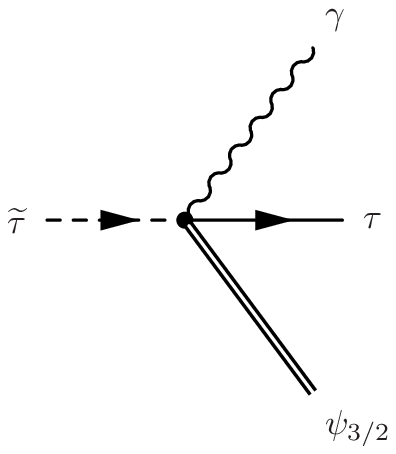}}}
 \hfil
 \subfigure[{}\label{fig:4particleInteractionM2R}]{
		\CenterObject{\includegraphics{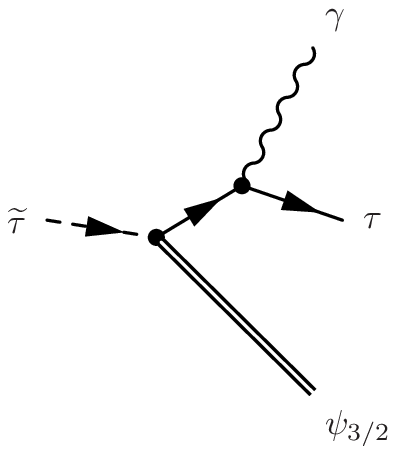}}}
 \hfil
  \subfigure[{}\label{fig:4particleInteractionM3R}]{
		\CenterObject{\includegraphics{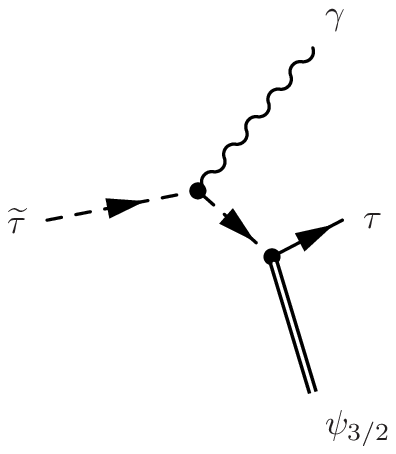}}}
\end{center}		
\caption{Diagrams contributing to $\stau\to\tau+\psi_{3/2}+\gamma$ at
tree level.  We do not take into account the diagram with a neutralino
intermediate state. It turns out that (a) is the crucial ingredient to
prove the spin-3/2 nature of the gravitino.}
\label{fig:4particleInteraction}
\end{figure}

To this end, we consider the 3-body decay
$\stau_\mathrm{R}\to\tau_\mathrm{R}+\psi_{3/2}+\gamma$,
leaving final states with $W$- or $Z$-bosons for future studies.  We
only consider the diagrams of Fig.~\ref{fig:4particleInteraction} and
restrict ourselves to a pure `right-handed' NSP $\stau_\mathrm{R}$. 
Here, we have neglected diagrams with neutralino intermediate states, assuming 
that they are suppressed by a large neutralino mass. 

In order to prove the spin-3/2 nature of the invisible particle, we
compare the 3-body decay with final state gravitino with the corresponding 
decay involving a hypothetical neutralino $\lambda$. As an example, we
consider the Yukawa coupling\footnote{This interaction would arise from gauging
the anomaly free U(1) symmetry $L_{\tau}-L_{\mu}$, the difference of $\tau$- and
$\mu$-number, in the MSSM, with $\lambda$ being the gaugino.},
\begin{equation}
 \mathscr{L}_\mathrm{Yukawa}\,=\,
  h\left(\stau_\mathrm{R}^*\,\overline{\lambda}\,P_\mathrm{R}\,\tau
         +\stau_\mathrm{L}^*\,\overline{\lambda}\,P_\mathrm{L}\,\tau \right)
+\text{h.c.}\;.
\label{eq:LYukawa}
\end{equation}
Accidentally, the coupling $h$ could be very small, such that the $\stau$ decay
rate would be consistent with the rate given in Eq.~(\ref{eq:2bodyDecayRateWithTau}).

Also the goldstino has Yukawa couplings of the type given in Eq.~(\ref{eq:LYukawa}).
The full interaction lagrangian is obtained by performing the substitution 
$\psi_\mu\to \sqrt{{2\over 3}}{1\over m_{3/2}}\partial_\mu\chi$ in the
supergravity lagrangian.
Using the equations of motion one finds for the non-derivative form of the
effective lagrangian \cite{Lee:1998aw},
\begin{equation}
 \mathscr{L}_\mathrm{eff}
 \,=\,
 \frac{m_{\widetilde{\tau}}^2}{\sqrt{3}M_\mathrm{P}\,m_{3/2}}
 \left(\widetilde{\tau}_\mathrm{R}^*\,\overline{\chi}\,P_\mathrm{R}\,\tau
 +\widetilde{\tau}_\mathrm{R}\,\overline{\tau}\,P_\mathrm{L}\,\chi\right)
 -\frac{m_{\widetilde{\gamma}}}{4\sqrt{6}M_\mathrm{P}\,m_{3/2}}
 \overline{\chi}[\gamma^\mu,\gamma^\nu]\,\widetilde{\gamma}\,F_{\mu\nu}\;,
 \label{eq:GoldstinoLagrangian}
\end{equation}
where we have neglected a quartic interaction term which is irrelevant for our
discussion. Note that the goldstino coupling to the photon supermultiplet is 
proportional to the photino mass $m_{\widetilde{\gamma}}$. As a consequence, 
the contribution to $\stau$-decay with intermediate photino is not suppressed 
for very large photino masses. As we shall see, this leads to significant 
differences between the angular distributions for pure Yukawa and goldstino 
couplings.

\begin{figure}[!h]
\begin{center}
 \subfigure[Kinematical configuration. The arrows denote the momenta.
 \label{fig:KinematicalSetupA}]{
 	\CenterObject{\includegraphics{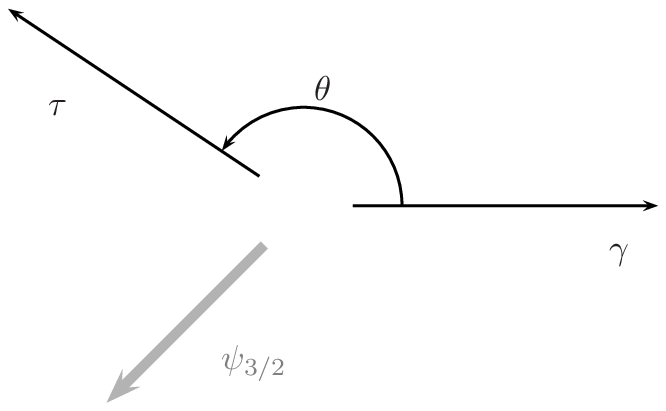}}}
 \qquad
 \subfigure[Characteristic spin-$3/2$ process. The thick arrows represent the
 	spins.\label{fig:TypicalSpin32}]{
 	\CenterObject{\includegraphics{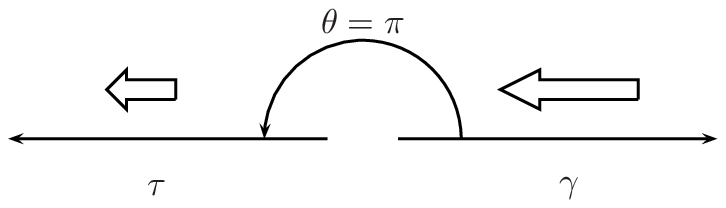}}}
\end{center}	
 \caption{(a) shows the kinematical configuration of the 3-body decay. 
  (b) illustrates the characteristic spin-3/2 process:
 if photon and $\tau$-lepton move in opposite directions and the spins 
 add up to $3/2$, the invisible particle also has spin $3/2$.}
 \label{fig:KinematicalSetup}
\end{figure}

In the following we discuss two methods to determine the spin of the 
invisible particle. The first one is based on a double differential angular 
and energy distribution, the second one makes use of the angular distribution 
of polarized photons.

In $\stau$-decay both, photon and $\tau$-lepton will mostly be very energetic.
Hence the photon energy $E_{\gamma}$ and the angle $\theta$ between $\tau$
and $\gamma$ can both be well measured (cf.~Fig.~\ref{fig:KinematicalSetupA}). 
We can then compare the differential decay rate
\begin{equation}
 \Delta(E_\gamma ,\cos\theta)\,=\,{1\over\alpha\,\Gamma_{\widetilde{\tau}}}
 \frac{\D^2\Gamma(\widetilde{\tau}\to\tau+\gamma+X)}{\D E_\gamma\,\D \cos\theta}\;,
\end{equation}
for the gravitino LSP ($X=\psi_{3/2}$) and the hypothetical neutralino 
($X=\lambda$) with pure Yukawa coupling. Details of the calculation are 
given in appendix~\ref{app:Details}.

\begin{figure}[t]
\begin{center}
 \subfigure[Gravitino]{
 	\CenterObject{\includegraphics[scale=0.85]{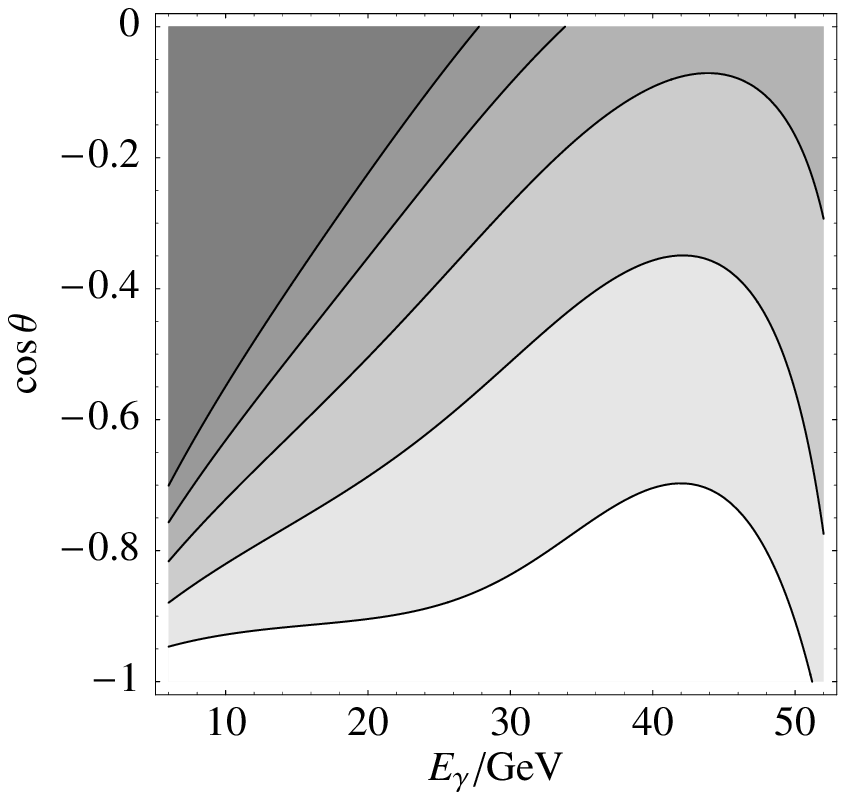}}}
 \quad
 \subfigure[Spin-1/2 neutralino]{
 	\CenterObject{\includegraphics[scale=0.85]{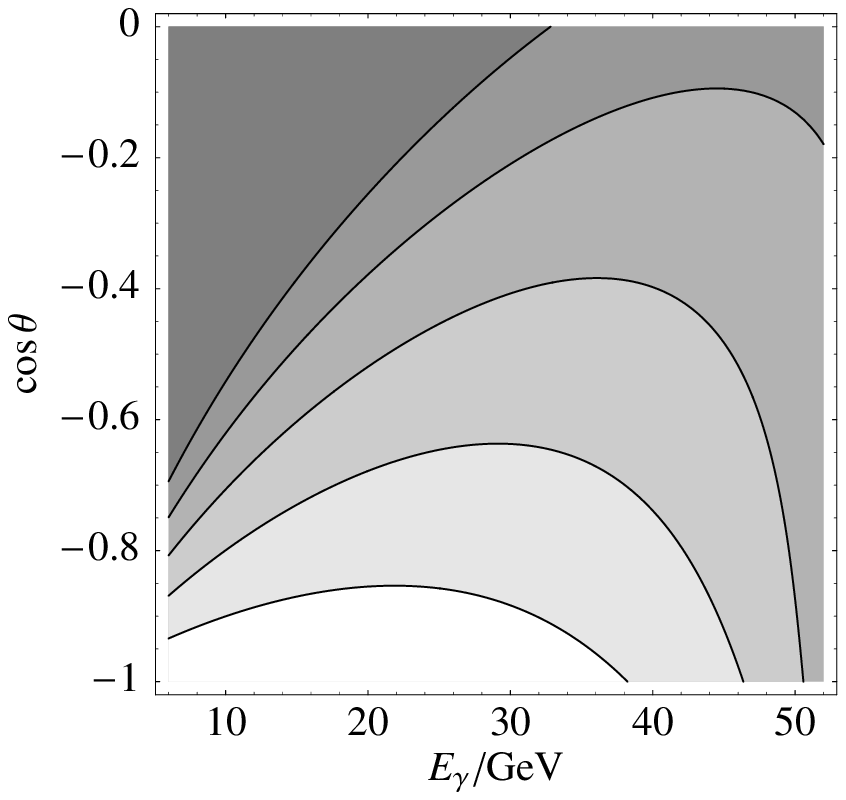}}}
\end{center}
\caption{Contour plots of the differential decay rates for
 (a) gravitino $\psi_{3/2}$ and (b) neutralino $\lambda$.
 $m_{\widetilde{\tau}}=150\,\mathrm{GeV}$ and 
 $m_{X}=75\,\mathrm{GeV}$ ($X=\psi_{3/2},\lambda$).
 The boundaries of the different gray shaded regions (from bottom to top)
 correspond to $\Delta(E_\gamma ,\cos\theta)[\mathrm{GeV}^{-1}]
 =10^{-3}, 2\times10^{-3}, 3\times10^{-3}, 4\times10^{-3}, 5\times10^{-3}$.
 Darker shading implies larger rate.}
\label{fig:CompareDifferentialDecayRate}
\end{figure}

In the forward direction, $\cos\theta \geq 0$, bremsstrahlung 
(cf.~Fig.~\ref{fig:4particleInteractionM2R}) dominates, and final states with 
gravitino and neutralino look very similar.
In the backward direction, $\cos\theta < 0$, the direct coupling 
Fig.~\ref{fig:4particleInteractionM1R} is
important, and the angular and energy distribution differs significantly for
gravitino and neutralino. This is demonstrated by 
Fig.~\ref{fig:CompareDifferentialDecayRate} where 
$m_{\widetilde{\tau}}=150\,\mathrm{GeV}$ and 
$m_{3/2}=75\,\mathrm{GeV}$ ($m_{\lambda}=75\,\mathrm{GeV}$). The two differential
distributions are qualitatively different and should allow to distinguish
experimentally gravitino and neutralino.

Note that even for very small masses $m_{3/2}$ and $m_\lambda$, the differential
decay rates $\Delta$ for gravitino $\psi_{3/2}$ and neutralino $\lambda$ are 
distinguishable. In this small mass limit, $\psi_{3/2}$ can effectively be 
described by the goldstino $\chi$ (with the interaction 
\eqref{eq:GoldstinoLagrangian}), and the differential decay rates for 
$\psi_{3/2}$ and $\chi$ essentially coincide. The discrepancy 
between $\lambda$ and $\chi$ stems from the additional photino contribution, 
as discussed below \eqref{eq:GoldstinoLagrangian}.
This makes it possible to discriminate the goldstino from the neutralino
even for very small masses.

The second method to test the spin-3/2 nature is intuitively more straightforward 
though experimentally even more challenging than the first one. The main point is
obvious from Fig.~\ref{fig:TypicalSpin32} where a left-handed photon and a
right-handed $\tau$ move in opposite directions. Clearly, this configuration is
allowed for an invisible spin-3/2 gravitino but forbidden for spin-1/2 neutralino.
Unfortunately, measuring the polarizations is a difficult task. 

As Fig.~\ref{fig:TypicalSpin32} illustrates, the spin of the invisible particle
influences the angular distribution of final states with polarized photons and
$\tau$-leptons. Again the difference between gravitino and neutralino is due to 
the direct coupling shown in Fig.~1(a) and most significant in the backward direction.
An appropriate observable is the angular asymmetry
\begin{equation}
 A_\mathrm{RL}(\cos\theta)\,=\,
 \frac{\displaystyle 
 \frac{\D \Gamma}{\D\cos\theta}(\stau_\mathrm{R}
 \to\tau_\mathrm{R}+\gamma_\mathrm{R}+X)
 -\frac{\D \Gamma}{\D\cos\theta}(\stau_\mathrm{R}
 \to\tau_\mathrm{R}+\gamma_\mathrm{L}+X)}{\displaystyle
 \frac{\D \Gamma}{\D\cos\theta}(\stau_\mathrm{R}
 \to\tau_\mathrm{R}+\gamma_\mathrm{R}+X)
 +\frac{\D \Gamma}{\D\cos\theta}(\stau_\mathrm{L}
 \to\tau_\mathrm{R}+\gamma_\mathrm{L}+X)}\;,
\end{equation}
where $X$ denotes gravitino ($X = \psi_{3/2}$) and neutralino 
($X=\lambda$). Here, we also study the angular asymmetry for a pseudo-goldstino
in the final state ($X=\chi$). Like a pseudo-Goldstone boson, the
pseudo-goldstino has goldstino couplings and a mass which explicitly breaks
global supersymmetry. Notice that, as mentioned above, the photino does not
decouple in this case.

\begin{figure}[!h]
\begin{center}
 \subfigure[$m_X=1\,\mathrm{GeV}$.]{
 	\CenterObject{\includegraphics[scale=0.85]{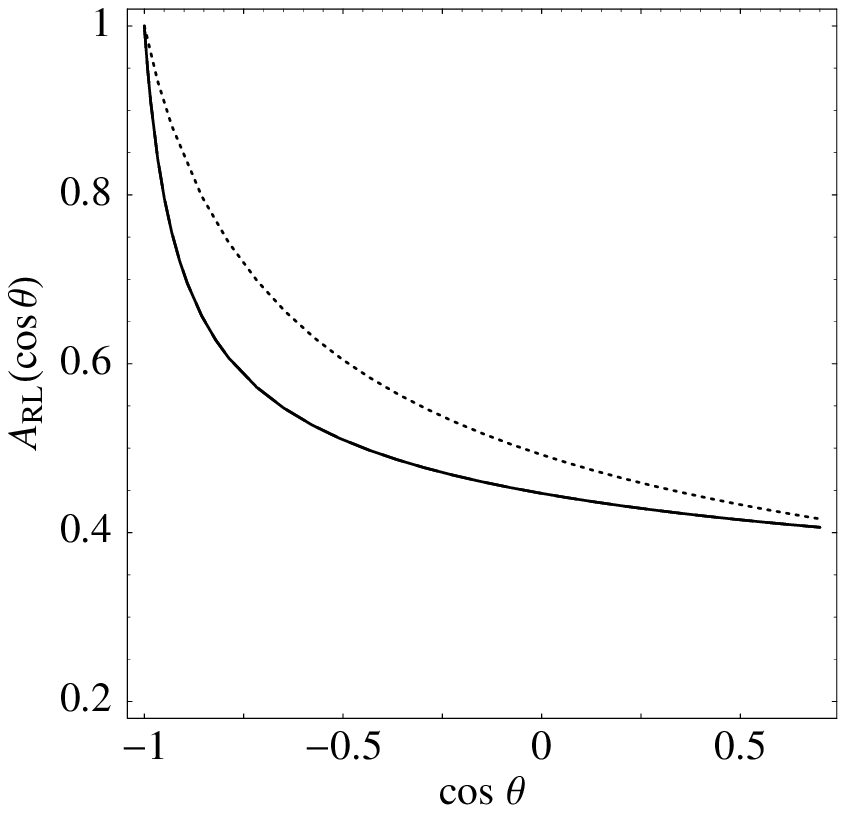}}}
 \hfil
 \subfigure[$m_X=10\,\mathrm{GeV}$.]{
 	\CenterObject{\includegraphics[scale=0.85]{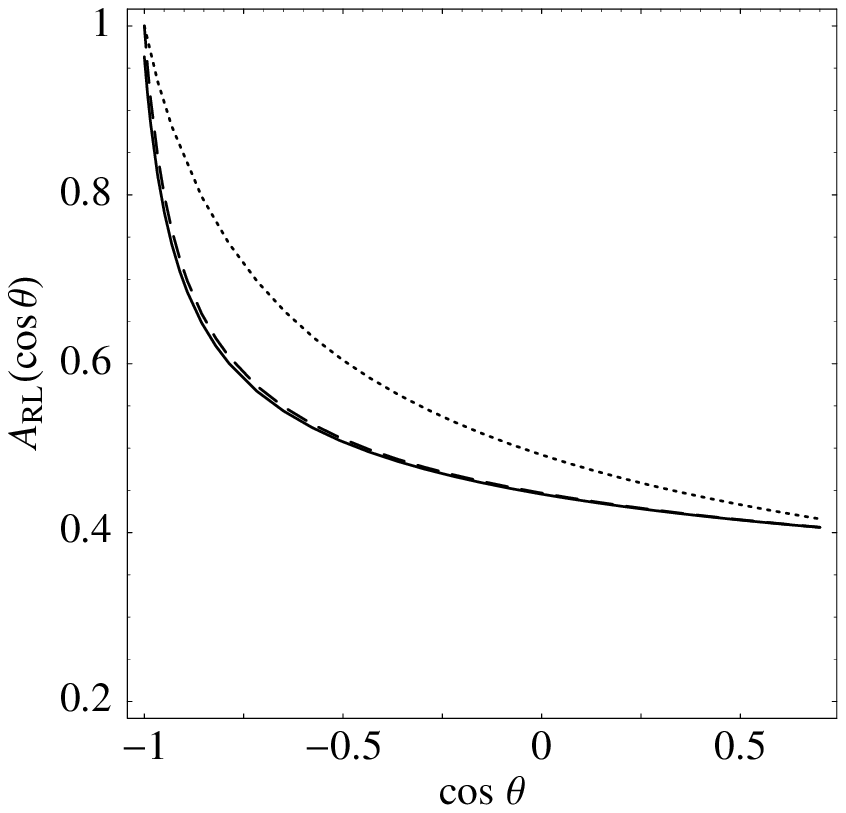}}}
 \\
 \subfigure[$m_X=30\,\mathrm{GeV}$.]{
 	\CenterObject{\includegraphics[scale=0.85]{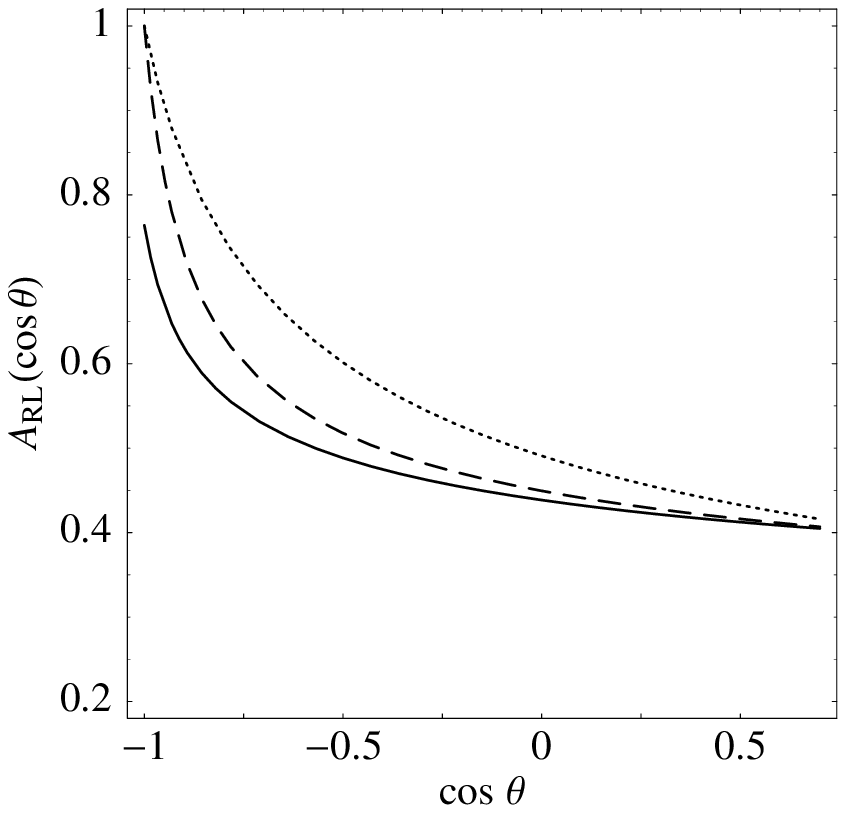}}}
 \hfil
 \subfigure[$m_X=75\,\mathrm{GeV}$.]{
 	\CenterObject{\includegraphics[scale=0.85]{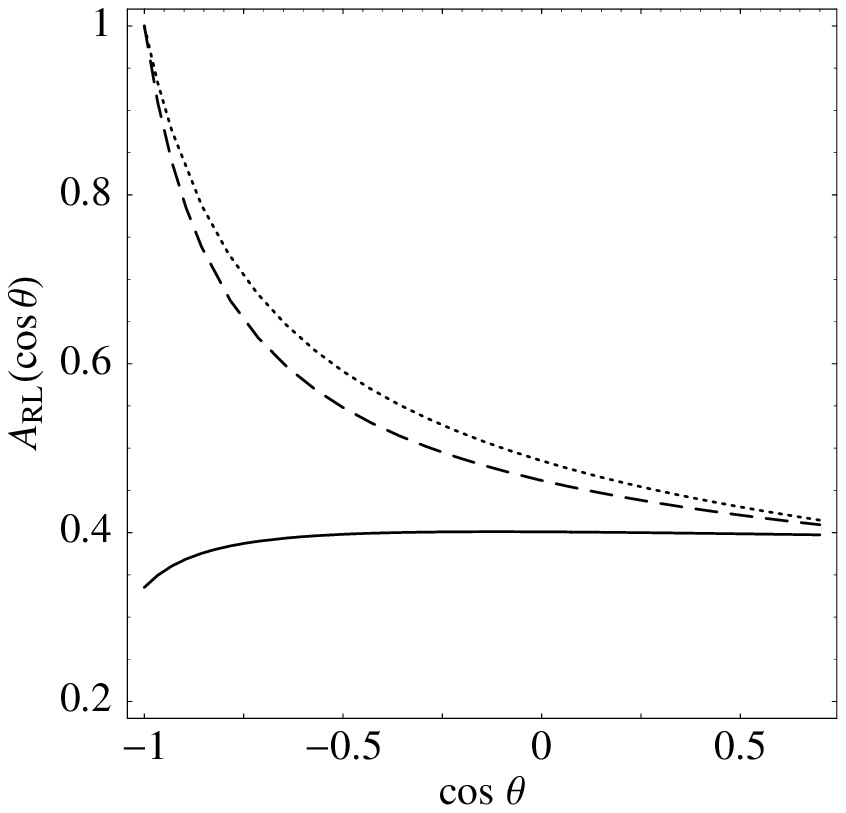}}}
\end{center} 
\caption{Angular asymmetries for gravitino $\psi_{3/2}$ (solid curve), goldstino
$\chi$ (dashed curve) and neutralino $\lambda$ (dotted curve).
We use $m_{\stau}=150\,\mathrm{GeV}$ and cut the photon energy
below $10\%$ of the maximal photon energy (cf.\ App.~\ref{app:Details}).  
Note that the asymmetries only depend on the ratio $r=m_X^2/m_{\stau}^2$
($X=\psi_{3/2}, \chi,\lambda$).}
\label{fig:CompareLandR}
\end{figure}

The three angular asymmetries are shown in Fig.~\ref{fig:CompareLandR} for 
$m_{\stau}=150\,\mathrm{GeV}$ and different masses of the invisible particle.
As expected, the decay into right-handed $\tau$ and left-handed photon
at $\theta=\pi$ is forbidden for spin-1/2 invisible particles ($\chi$ and
$\lambda$), whereas it is allowed for the spin-3/2 gravitino. This is clearly
visible in Figs.~\ref{fig:CompareLandR}(c,d); for small gravitino masses
the goldstino component dominates the gravitino interaction as illustrated by
Figs.~\ref{fig:CompareLandR}(a,b).

Our discussion is easily generalized to the case where the NSP is a linear
combination of $\stau_\mathrm{R}$ and $\stau_\mathrm{L}$. One then needs
further information on the left-right mixing angle $\varphi_\tau$, which
could be provided by a direct measurement of the $\tau$-polarization 
or by the coupling to $W$-boson.

Let us finally comment on the experimental feasibility of the gravitino spin
determination. The angular distribution of the 3-body decay is peaked in forward
direction ($\theta=0$). 
Hence, a large number of events is needed for the spin measurement.
Compared to the 2-body decay, backward ($\cos\theta<0$) 3-body decays are 
suppressed by $\sim 10^{-1}\times\alpha\simeq10^{-3}$. Requiring 10\dots100 
events for a signal one therefore needs $10^4$ to $10^5$ $\stau$s, which 
appears possible at the LHC
and also at a Linear Collider according to the discussion in 
Sec.~\ref{sec:GravitinoMass}.

\subsection{Gravitino cosmology}\label{sec:Cosmology}

The existence of gravitinos imposes severe constraints on the early
history of our universe. If the gravitino is the LSP and stable, as
assumed in our investigation, there are two important constraints which
we shall briefly discuss in this section.  The first one arises for
large reheating temperatures $T_R$ after inflation,
which may lead to a relic gravitino abundance exceeding the observed
cold dark matter density. This `overclosure' constraint implies an upper
bound on the reheating temperature~\cite{Moroi:1993mb}. Note, however, that 
there are several mechanisms which avoid this constraint and which, in 
addition, explain the observed cold dark matter in terms of 
gravitinos~\cite{Bolz:1998ek}.

The second constraint concerns the decay of the long-lived NSP $\stau$.  
If it occurs during or after nucleosynthesis (BBN), it may spoil
the successful predictions of BBN~\cite{Moroi:1993mb,Fujii:2003nr}. A
recent detailed analysis \cite{Kohri:talk} shows that
the hadronic decay of a heavy particle during or after BBN imposes
severe constraints on its abundance and lifetime. If the 3-body decay
$\widetilde{\tau}\to \psi_{3/2} + \tau + Z$ is allowed, 
one finds the upper bound on the $\stau$-lifetime 
$(\Gamma_{\stau})^{-1}\lesssim 10^3\,\mathrm{s}$, or equivalently, 
$m_{3/2}\lesssim 0.4\,\mathrm{GeV}\,(m_{\stau}/150\,\mathrm{GeV})^{5/2}$. 
Note that in the case of non-zero left-right mixing $\varphi_\tau$, 
processes involving $W$ also have to be taken into account.
On the other hand, if hadronic $\stau$ decays are sufficiently suppressed, 
which is the case for $m_{\widetilde{\tau}}-m_{3/2} < m_Z$, only the effect 
of the electromagnetic NSP decay~\cite{Kawasaki:2000qr}
has to be taken into account. The allowed mass range is then extended to
$100\,\mathrm{GeV}\lesssim m_{\stau}\lesssim 130\,\mathrm{GeV}$ 
and $(m_{\stau}-m_Z)\lesssim m_{3/2}\lesssim 35\,\mathrm{GeV}$ for a
typical pair annihilation cross section $\sigma_{\stau}$ of $\stau$s, and
to $100\,\mathrm{GeV}\lesssim m_{\stau}\lesssim 350\,\mathrm{GeV}$ and 
$(m_{\stau}-m_Z)\lesssim m_{3/2}\lesssim 260\,\mathrm{GeV}$, if 
$\sigma_{\stau}$ is enhanced by a factor 100 \cite{Ibe:communications}. 
Note that for larger gravitino masses the spin determination is easier, 
as discussed in Sec.~\ref{sec:GravitinoSpin}.

Finally, we should mention that the above BBN constraints disappear if
there is sufficient entropy production between the decoupling of the NSP
at $T_d \simeq\mathcal{O}(10\,\mathrm{GeV})$ and BBN at 
$T_\mathrm{BBN}\simeq \mathcal{O}(\mathrm{MeV})$, or if the
reheating temperature $T_R$ is lower than $T_d$, $T_\mathrm{BBN}<T_R<T_d$, 
so that there is no thermal production of $\stau$s.

\subsection{Conclusions}\label{sec:Conclusions}

We have discussed how one may discover the massive gravitino, and thereby
supergravity, at the LHC or a future Linear Collider, if the gravitino is the
LSP and a charged slepton is the NSP. With the gravitino mass inferred from
kinematics, the measurement of the NSP lifetime will test an unequivocal
prediction of supergravity. For gravitino masses 
$m_{3/2}\gtrsim\mathcal{O}(\mathrm{GeV})$ also the determination of the
gravitino spin appears feasible.

\subsection*{Acknowledgements}

We would like to thank M.~Ibe, T.~Plehn, D.~Rainwater, P.~Schleper, C.~Youngman 
and P.~M.~Zerwas for valuable discussions.

\renewcommand{\thesection}{\Alph{section}}
\renewcommand{\thesubsection}{\Alph{subsection}}
\def\theequation{\Alph{subsection}.\arabic{equation}}
\renewcommand{\thetable}{\Alph{section}.\arabic{table}}
\setcounter{subsection}{0}
\setcounter{equation}{0}

\subsection{3-body slepton decays}\label{app:Details}

This appendix provides some details of the calculations leading to the
results of Sec.~\ref{sec:GravitinoSpin}. For simplicity we take
$m_\tau=0$. The diagrams of Fig.~\ref{fig:4particleInteraction} yield
\begin{subequations}
\label{eq:Gravitino-M2}
\begin{eqnarray}
  \lefteqn{\hspace*{-0.3cm}
  \sum\limits_{\mathrm{spins}}
  \left|\mathcal{M}(\stau_\mathrm{R}\to\tau_\mathrm{R}+\gamma_\mathrm{L}
  +\psi_{3/2})\right|^2
  \,=\,	
  \frac{2\,e^2\,m_{\stau}^2}{3\,M_\mathrm{P}^2}
  \left(1-r-2\eta\right)\frac{(1-r)^2(2-z)+12\,r\,z\,\eta^2}{4\,r\,z\,\eta^2}
  \;,}
  \nonumber\\ 
 \label{eq:Gravitino-left}
 \\
  \lefteqn{\hspace*{-0.3cm}
    \sum\limits_{\mathrm{spins}}
  \left|\mathcal{M}(\stau_\mathrm{R}\to\tau_\mathrm{R}+\gamma_\mathrm{R}
  +\psi_{3/2})\right|^2
  \,= \, 	
  \frac{2\,e^2\,m_{\stau}^2}{3\,M_\mathrm{P}^2}
  \left\{
  \frac{(1-r)^2(2-z)(1-r+2\,\eta)}{4\,r\,z\,\eta^2}
  \right.
  \;}\nonumber\\
  \lefteqn{
  \,\,\,{}
  +\frac{(1-r)^2[2-(1-r)\,z]}{r\,z\,(1-r-2\,\eta)}
  +\frac{12-10\,(1-r)\,z-2\,r\,(4-r)\,z^2+(1+r-2\,r^2)\,z^3}{r\,z^2(1-z\,\eta)}
  \;}\nonumber\\
  \lefteqn{
  \left.
  \,\,\,{}
  -\frac{[2-(1-r)z]^2}{r\,z^2\,(1-z\,\eta)^2}
  +\frac{[-8+6\,(1-r)\,z-(1-7r)\,z^2]-2\,\eta\,[2\,z-(1-4\,r)z^2]}{r\,z^2}
  \right\}
  \;,}\nonumber\\
  \label{eq:Gravitino-right}
\end{eqnarray}
\end{subequations}
where $r=m_{3/2}^2/m_{\stau}^2$, $\eta=E_\gamma/m_{\stau}$ and
$z=1-\cos\theta$.  The corresponding transition probabilities for the
(hypothetical) spin-1/2 particle $\lambda$ read (now with
$r=m_\lambda^2/m_{\stau}^2$)
\begin{subequations}
\begin{eqnarray}
 \lefteqn{\hspace*{-0.3cm}
 \sum\limits_{\mathrm{spins}}
 |\mathcal{M}(\widetilde{\tau}_\mathrm{R}\to \tau_\mathrm{R}
 +\gamma_\mathrm{L}+\lambda)|^2
 \,=\,
 e^2\,h^2\,
 \frac{(2-z)(1-r-2\,\eta)}{2\,z\,\eta^2}
 \; , } 
 \\
 \lefteqn{ \hspace*{-0.3cm}
 \sum\limits_{\mathrm{spins}}
 |\mathcal{M}(\widetilde{\tau}_\mathrm{R}\to \tau_\mathrm{R}
 	+\gamma_\mathrm{R}+\lambda)|^2
 \,=\,
 e^2\,h^2\,
 \frac{(1-r)^2(2-z)+4\,r\,z\,\eta^2}{2\,z\,\eta^2(1-r-2\,\eta)}
 \; . } 
\end{eqnarray}
\end{subequations}
In the case of pseudo-goldstino $\chi$ with interactions described 
by Eq.~\eqref{eq:GoldstinoLagrangian} one has to include the diagram with
the photino intermediate state. We then obtain in the limit of a large 
photino mass $m_{\widetilde{\gamma}}$ (with $r=m_\chi^2/m_{\stau}^2$)
\begin{subequations}
  \label{eq:Goldstino-M2}
\begin{eqnarray}
 \lefteqn{\hspace*{-0.3cm}
 \sum\limits_{\mathrm{spins}}
 |\mathcal{M}(\widetilde{\tau}_\mathrm{R}\to \tau_\mathrm{R}
 +\gamma_\mathrm{L}+\chi)|^2
 \,=\,
 e^2\,\frac{m_{\stau}^4}{3\,m_{3/2}^2\,M_\mathrm{P}^2}\,
 \frac{(2-z)(1-r-2\,\eta)}{2\,z\,\eta^2}
 \; , } 
 \label{eq:Goldstino-left}
 \\
 \lefteqn{\hspace*{-0.3cm}
 \sum\limits_{\mathrm{spins}}
 |\mathcal{M}(\widetilde{\tau}_\mathrm{R}\to \tau_\mathrm{R}
 	+\gamma_\mathrm{R}+\chi)|^2
 \,=\,
 e^2\,\frac{m_{\stau}^4}{3\,m_{3/2}^2\,M_\mathrm{P}^2}\,
 \left\{
 \frac{(1-r)^2(2-z)+4\,r\,z\,\eta^2}{2\,z\,\eta^2(1-r-2\,\eta)}
 \right.
 \;  } 
 \nonumber\\
 \lefteqn{
 \,\,\,{}
 \left.
 +\frac{2\,z\,\eta^2(1-r-2\,\eta)[2-(1-r)\,z]}{(1-z\,\eta)^2}
 +\frac{2\,z\,(1-r-2\,\eta)}{1-z\,\eta}
 -4\,(1-r)
 \right\}
 \; . }
 \label{eq:Goldstino-right}
\end{eqnarray}
\end{subequations}
The limit $r\to 0$ in Eqs.~\eqref{eq:Goldstino-M2} yields the results
for massless goldstino. Indeed, as can be seen by straightforward
calculation, they precisely reproduce the massless limits of gravitino
transition probabilities, which are obtained by taking the limit $r\to
0$ in Eqs.~\eqref{eq:Gravitino-M2} while keeping the SUSY breaking
parameter $\sqrt{3}m_{3/2}M_\mathrm{P}$ finite.

In order to present the angular distribution, we perform an $E_\tau$
integration,
\begin{equation}
 \frac{\D\Gamma}{\D \cos\theta}
 \,=\,\frac{1}{128\,\pi^3}\,
 \int\limits_{0}^{E_\tau^\mathrm{max}}\!\D E_\tau\,
 \frac{E_\tau\,\left[  m_{\stau}\,
         \left( m_{\stau} - 2\,E_\tau \right) - m_{X}^2 \right]}{m_{\stau}
		\,\left[ m_{\stau} - (1-\cos\theta)\,E_\tau  \right]^2 }
 \,\left|\mathcal{M}\right|^2\;,
\end{equation}
where $m_X$ denotes the mass of the invisible particle.  Notice that
the $|\mathcal{M}|^2$s have singularities coming from soft photon
($1/E_\gamma\propto 1/\eta$), soft $\tau$ ($1/E_\tau\propto
1/[1-r-2\eta]$) and a collinear divergence ($1/[1-\cos\theta] =
1/z$). The last two are not really divergent for finite $m_\tau$.  We
remove the soft photons from the rates. This procedure is justified by
the limited resolution of real detectors.  The requirement that
$E_\gamma\ge \delta\,E_\mathrm{max}$ (where
$E_\mathrm{max}=m_{\stau}(1-r)/2$ is the maximal energy of the photon)
leads to
\begin{equation}
 E_\tau\,\le\,
 E_\tau^\mathrm{max}(\delta)
 \,=\,
 \frac
     {m_{\stau}\,\left( m_{\stau}^2 - m_{X}^2\right)
       \,\left( 1 - \delta \right) }
     {2 m_{\stau}^2 - \delta\,
       \left(m_{\stau}^2-m_X^2\right)\,
       \left(1-\cos\theta\right)}
     \;.
\end{equation}
In Sec.~\ref{sec:GravitinoSpin}, $\delta=0.1$ is used, i.e.\ the
photons are cut below 10\% of their maximal energy.


\begin{thebibliography}{10}

\bibitem{Freedman:1976xh}
D.~Z. Freedman, P.~van Nieuwenhuizen, S.~Ferrara, Phys. Rev. \textbf{D13}
  (1976) 3214;
\\
S.~Deser, B.~Zumino, Phys. Lett. \textbf{B62} (1976) 335.

\bibitem{Kaplan:1999ac}
D.~E. Kaplan, G.~D. Kribs, M.~Schmaltz, Phys. Rev. \textbf{D62} (2000)
  035010;
\\
Z.~Chacko, M.~A. Luty, A.~E. Nelson, E.~Pont{\'o}n, JHEP \textbf{01}
  (2000) 003;
\\
K.~Inoue, M.~Kawasaki, M.~Yamaguchi, T.~Yanagida,
Phys. Rev. \textbf{D45} (1992) 328.

\bibitem{Nilles:1984ge}
H.~P. Nilles, Phys. Rept. \textbf{110} (1984) 1.

\bibitem{Giudice:1998bp}
G.~F. Giudice,  R.~Rattazzi, Phys. Rept. \textbf{322} (1999) 419.

\bibitem{Drees:1990yw}
M.~Drees, X.~Tata,
Phys. Lett. \textbf{B252} (1990) 695.

\bibitem{Beenakker:1997ch}
W.~Beenakker, R.~Hopker, M.~Spira, P.~M. Zerwas, Nucl. Phys. \textbf{B492}
  (1997) 51.

\bibitem{Maki:1998ih}
K.~Maki, S.~Orito, Phys. Rev. \textbf{D57} (1998) 554.

\bibitem{Aguilar-Saavedra:2001rg}
TESLA, Technical Design Report Part III,
eds.~R.-D.~Heuer, D.~Miller, F.~Richard, P.~Zerwas, 
DESY 01-011 (2001), hep-ph/0106315;\\
ACFA LC Working Group, K.~Abe {\it et al.}, KEK-REPORT-2001-11, 
hep-ph/0109166;\\
American LC Working Group, T.~Abe {\it et al.}, 
SLAC-R-570 (2001), hep-ex/0106055.

\bibitem{Feng:1997zr}
J.~L.~Feng, T.~Moroi,
Phys. Rev. \textbf{D58} (1998) 035001;\\
S.~P. Martin, J.~D. Wells, Phys. Rev. \textbf{D59} (1999) 035008.

\bibitem{Ambrosanio:2000ik}
S.~Ambrosanio, B.~Mele, S.~Petrarca, G.~Polesello,  A.~Rimoldi, JHEP
  \textbf{01} (2001) 014.

\bibitem{Ambrosanio:1999iu}
S.~Ambrosanio, G.~A. Blair, Eur. Phys. J. \textbf{C12} (2000) 287;
\\
P.~G.~Mercadante, J.~K.~Mizukoshi, H.~Yamamoto,
Phys. Rev. \textbf{D64} (2001) 015005.

\bibitem{Wess:1992cp}
J.~Wess, J.~Bagger, \emph{Supersymmetry and Supergravity}, Princeton
University Press, Princeton, New Jersey, 1992.

\bibitem{Hagiwara:2002fs}
Particle Data Group, K.~Hagiwara et~al., Phys. Rev. \textbf{D66} (2002)
  010001.

\bibitem{Lee:1998aw}
T.~Lee, G.-H. Wu, Phys. Lett. \textbf{B447} (1999) 83.

\bibitem{Moroi:1993mb}
T.~Moroi, H.~Murayama,  M.~Yamaguchi, Phys. Lett. \textbf{B303} (1993)
  289.

\bibitem{Bolz:1998ek}
M.~Bolz, W.~Buchm{\"u}ller, M.~Pl{\"u}macher, Phys. Lett. \textbf{B443}
  (1998) 209;
\\
M.~Fujii, T.~Yanagida, Phys. Lett. \textbf{B549} (2002) 273;
\\
W.~Buchm{\"u}ller, K.~Hamaguchi, M.~Ratz, Phys. Lett. \textbf{B574} (2003)
  156;
\\
M.~Fujii, M.~Ibe, T.~Yanagida, Phys. Rev. \textbf{D69} (2004)
015006.

\bibitem{Fujii:2003nr}
J.~L.~Feng, A.~Rajaraman, F.~Takayama,
Phys. Rev. \textbf{D68} (2003) 063504;\\
M.~Fujii, M.~Ibe, T.~Yanagida, Phys. Lett. \textbf{B579} (2004) 6.

\bibitem{Kohri:talk}
K.~Kohri, \emph{Talk given at 2003 {A}utumn {M}eeting of the {P}hysical
  {S}ociety of {J}apan}, 2003.

\bibitem{Kawasaki:2000qr}
M.~Kawasaki, K.~Kohri, T.~Moroi, Phys. Rev. \textbf{D63} (2001) 103502;
\\
R.~H. Cyburt, J.~R. Ellis, B.~D. Fields, K.~A. Olive, Phys. Rev.
  \textbf{D67} (2003) 103521.

\bibitem{Ibe:communications}
M.~Ibe, \emph{private communication}, 2004.

\end{thebibliography}
\end{document}